\documentclass[iop]{emulateapj-rtx4} % read by: ACROREAD
\shortauthors{Sekanina}
\shorttitle{A Super-Kreutz Sungrazer System}
\slugcomment{Version \today }

\newcommand{\rzeme}{$r_{\mbox{\tiny \boldmath $\!\oplus$}}$}
\begin{document}
\title{On The Existence Of A Super-Kreutz System of Sungrazing Comets}
\author{Zdenek Sekanina}
\affil{La Canada Flintridge, California 91011, U.S.A.}
\email{ZdenSek@gmail.com.}

\begin{abstract} % maximum length = 1920 characters
In the context of a recently proposed contact-binary model of the Kreutz
system, all its members~are products of the process of cascading fragmentation
of the two lobes of the parent, \mbox{Aristotle's} comet of 372~BC.  This
process presumably began with the lobes' separation from each other near
aphelion.  However, not every object in a Kreutz-like orbit is a Kreutz sungrazer.
Any surviving sungrazer~that had split off from the progenitor {\it before\/}
the lobes separated, as well as {\it its\/} surviving fragments born in any
subsequent tidal or nontidal event, are by definition {\it not\/} members of
the Kreutz system.  Yet, as parts of the same progenitor, they belong\,---\,as
do all Kreutz sungrazers\,---\,to a broader assemblage~of related objects, which
I refer to as a {\it super-Kreutz system\/}.  After estimating the ratio of
the~number of super-Kreutz members to nonmembers among {\it potential historical
sungrazers\/}, I generate representative extended {\it pedigree charts\/} for
both the Kreutz system and super-Kreutz system.  While the~fragmentation paths
and relationships among the individual sungrazers or potential sungrazers in
the~two~charts are (with at most a few exceptions) arbitrary, the purpose of
the exercise is to suggest that~the~Kreutz system proper could in effect
represent an ultimate deagglomeration stage of the super-Kreutz system.
% is organic part of a broader, more massive
% fragmentation process of sungrazing~comets.
\end{abstract}

\section{Introduction}
%%% Sec. 1
I recently introduced a new, contact-binary model for the Kreutz system of
sungrazing comets (Sekanina 2021, hereafter referred to as Paper~1; Sekanina
2022a, referred to as Paper~2).  The model incorporated novel features
described in greater detail in separate papers (Sekanina 2022b~=~Paper~3;
Sekanina \& Kracht 2022~=~Paper~4) and its feasibility was demonstrated by
computer simulating the orbital motions of the progenitor's major fragments
(Paper~4).  Likewise addressed were the age of the Kreutz system, the
populations of sungrazers that made it up, effects of cascading
fragmentation, and properties of the major, naked-eye members on the one
hand and the dwarf members on the other hand (Papers~\mbox{1--4} and
Sekanina 2022c~=~Paper~5).

By\,its\,very\,nature~the~model~implies\,that~\mbox{fragmentation}
proceeds along the entire orbit about the Sun.  Depending on the location,
the fragmentation is triggered either {\it tidally\/} (near perihelion),
in which case the primarily perturbed element is the orbital period;
or {\it nontidally\/} (far from the Sun), when greatly perturbed are
the perihelion distance and the angular elements.  As a result, the
distribution of Kreutz sungrazers along the orbit is the product of a
mix of stochastic and pseudo-periodic processes.  The fragments pass
through perihelion in principle randomly, except that sometimes they
do so in clusters, which could be separated from one another
quasi-uniformly.

Because of the cascading nature of fragmentation, the early number of
fragments increases --- and their dimensions decrease --- exponentially
with time.  Granted that the progenitor --- presumed in the contact-binary
model to be Aristotle's comet of 372~BC --- was of enormous dimensions,
perhaps 100~km or more in diameter, its many fragments are still of impressive
sizes, displaying brilliant tails near perihelion.  Yet, because of the
extremely hostile environment near the Sun, the lifetime of sungrazers
is believed to be rather short.

The model's fundamental trait is the {\it initial\/} fragmentation event
--- a low-velocity separation of the progenitor's two lobes from one
another --- taking place near aphelion around 5~BC.  Without this episode
I find it impossible to understand the major orbital gap --- in the
longitude of the ascending node, other angular elements, and perihelion
distance --- between the largest presumed surviving fragments of the two
lobes, the Great March Comet of 1843 (C/1843~D1) and the Great September
Comet of 1882 (C/1882~R1), respectively.

It may appear implausible to the reader that the {\it first\/} fragmentation
event occurred at aphelion, when the progenitor was subjected to powerful
disruptive forces at perihelion.  One has to keep in mind that the
episode~was, {\it by definition\/}, the first only in the history of
the Kreutz system.  Aristotle's comet had surely fragmented numerous
times, especially in close proximity of perihelion, long {\it before\/}
the bond between the two lobes of the contact binary failed.  Only one
of the many products of this early fragmentation would eventually give
birth to the Kreutz system.  All others could {\it never\/} have become
its members.  Yet,
% The products of this early fragmentation, moving in sungrazing orbits
% similar (except for the orbital periods) to the progenitor's orbit,
% could {\it never\/} have been members of the Kreutz system. Yet,
given their common ancestor, the members and nonmembers of the Kreutz system
proper alike do make up a greater family of closely related objects that
I refer to below as a {\it super-Kreutz system\/}.

To document the existence of this broader assemblage of sungrazers, I
compare a model-based, representative extended pedigree chart of the
super-Kreutz system with that of the Kreutz system proper,
subject~to~limited constraints provided by the definite Kreutz comets and their
fragmentation features, potential historical~sun\-grazers, and other
evidence noted in the following. \\[0.1cm]

\section{Kreutz System and Constraints\\on Its Pedigree Chart}
{\vspace{-0.065cm}}
Next to Aristotle's comet as progenitor, the contact-binary model recognizes
the daylight comets of AD~363, noted by Ammianus Marcellinus in his {\it
Res Gestae\/}~(Rolfe 1940), as the second-generation objects (i.e., the
first-generation fragments) of the Kreutz system.  Other sungrazers in
progenitor-like orbits and at perihelion in the centuries before or after
AD~363 could be members~of~the super-Kreutz system but not the Kreutz
system.

The Kreutz objects of the third generation can at best be recognized
with a limited degree of confidence in the cases in which the orbital
periods of the respective objects of the fourth generation have been
determined~fairly accurately, a condition satisfied nowadays by five
Kreutz sungrazers only.  Next to the Great March Comet~of~1843 and
the Great September Comet of 1882, they are Ikeya-Seki (designated as
C/1965~S1), Pereyra (C/1963~R1), and Lovejoy (C/2011~W3).

Major new information on the third-generation objects of the Kreutz
system was offered in Papers~2 and 4.  It was concluded in Paper~4
that X/1106~C1, the widely-observed probable sungrazer, could not be
the parent to comet Ikeya-Seki, contrary to what has often been presented
in the literature.  Linkage of the motion of the single nucleus of
Ikeya-Seki before perihelion with the motion of the principal fragment~A
over different periods of time after perihelion suggests that previously
the comet was at perihelion around AD~1140, with an uncertainty of
a few years.  An apparent, formerly unrecognized sungrazer first
observed in early September 1138 (Ho 1962), having passed perihelion
around August~1, appears to be the only plausible candidate.  Given
that C/1882~R1 and C/1965~S1 had been one comet up to the 12th
century perihelion (Marsden 1967), the above result is in remarkable
agreement with Marsden's back integration of the motion of the
principal fragment B [No.\,2 in Kreutz's (1891) notation] of C/1882~R1,
for which he derived a perihelion time in April 1138!

In contrast, strong evidence suggests that X/1106~C1~is the parent to
the sungrazer C/1843~D1 and Population~I associated with it.  Indeed,
the four generations of sungrazers in 372~BC, AD~363, 1106, and 1843
make up an almost perfect progression, implying a near-constant period
of 738~yr, compatible with the observations of the Great March Comet
of 1843 (Sekanina \& Chodas 2008).\,\,\,

Another candidate for a third-generation Kreutz~member presented in
Papers~2 and 4 was the Chinese-Korean comet first seen in September
1041 (Hasegawa \& Nakano 2001; England 2002), the proposed parent to
Pereyra; it passed perihelion in early August according to
Hasegawa \& Nakano.

To pinpoint a parent to the sungrazer Lovejoy is more difficult.
The comet's orbit determination by Sekanina \& Chodas (2012) showed
that a near-perihelion osculating value of the orbital period was
698~yr, equivalent to a barycentric orbital period of 683~yr, with
an uncertainty of $\pm$2~yr.  The previous perihelion would have
occurred in AD~1329, when no comet, least of all a sungrazer, was
under observation (Ho 1962; Hasegawa 1980).

The complication with the orbital period of Lovejoy was the
disintegration of the nucleus approximately 40~hr after perihelion
(Sekanina \& Chodas 2012).  From that time on, the sunward tip of
Lovejoy's headless condensation was made up of the largest dust
particles that survived intact.  Subjected to solar radiation
pressure, the headless comet began to move in a gravity field
that was slightly weaker than before the nucleus demise.  The
antisolar radial shift was taken into account in the orbit
determination and the (essentially constant) normalized magnitude
of the effect of solar radiation pressure was derived to equal
\mbox{$\beta_{\rm tip} = 0.00191 \pm 0.00042$} the solar
gravitational acceleration.

One could argue that solar radiation pressure triggers a perturbation
of the radial component of the comet's orbital acceleration, whose
magnitude at 1~AU from the Sun is $\Gamma$ and whose integrated
effect on the orbital period $P$ between times $t_0$ and $t$ is
$\Delta P_{\rm rp}$, for which perturbation theory gives
\begin{equation}
\frac{\Delta P_{\rm rp}}{P} = \frac{3e \sqrt{p}}{k (1 \!-\! e^2)} \,
  \Gamma \!\! \int_{t_0}^{t} \!\! g(r) \sin u \, dt,  % (1)
\end{equation}
where $k$ is the Gaussian gravitational constant{\vspace{-0.09cm}} (in
units of \mbox{AU$^\frac{3}{2}$\,day$^{-1}$}), $e$ is the orbit eccentricity,
\mbox{$p = q (1 \!+\! e)$}, $q$ is the perihelion distance (in AU),
$g(r)$ is the perturbation's dimensionless law of variation with
heliocentric distance \mbox{$r = r(t)$} (in AU) subject to a condition
\mbox{$g(1) = 1$}, $u$ is a true anomaly at time $t$ expressed in days.
Since the solar gravitational acceleration equals 0.593~cm~s$^{-2}$ at
1~AU from the Sun, the solar radiation pressure acceleration \mbox{$\Gamma
= 2.96 \:\!\!\times\! 10^{-4} \beta_{\rm tip}$}~AU~day$^{-2}$ and \mbox{$g(r)
=(\mbox{\rzeme}/r)^2$}, where \mbox{\rzeme\,\,=\,\,1 AU}.  From Kepler's
second{\vspace{-0.04cm}} law I substitute in Equation~(1) \mbox{$dt =
(r^2\!/k \sqrt{p}) \, du$}, take a true anomaly at the time of
disintegration, $u_{\rm d}$, to match the lower limit, $t_0$, and a
true anomaly at aphelion, \mbox{$u = \pi$}, to match
the upper limit, $t$, so that with $\Delta P_{\rm rp}$ and $P$ in years,
\begin{eqnarray}
\Delta P_{\rm rp} & = & 8.88 \:\!\! \times \! 10^{-4} \beta_{\rm tip} \, P \,
 \frac{e}{1 \!-\! e^2} \left( \! \frac{\mbox{\rzeme}}{k} \!
 \right)^{\!2} \! (1 \!+\! \cos u_{\rm d}) \nonumber \\[0.2cm]
 & \simeq & 1.74 \:\!\! \times \! 10^{-5} \beta_{\rm tip} \,
	\frac{P^{\frac{5}{3}}}{k^2 r_{\rm d}},  % (2)
\end{eqnarray}
where I introduce a heliocentric distance, $r_{\rm d}$, at the~time
of disintegration by approximating{\vspace{-0.055cm}} \mbox{$1 \!+\!
\cos u_{\rm d} = p/r_{\rm d}$} in the second line.  I also use the
equality of \mbox{$p/(1 \!-\! e^2) = P^{\frac{2}{3}}$}, approximate
\mbox{$e = 1$} in the numerator, and leave out \rzeme, which equals
unity.

Inserting for comet Lovejoy \mbox{$P \!=\! 683$ yr}, \mbox{$r_{\rm d}
\!=\! 0.144$ AU}, and the above value of $\beta_{\rm tip}$, I get
\begin{equation}
\Delta P_{\rm rp} = 41 \pm 9 \; {\rm yr},
\end{equation}
which implies that comet Lovejoy's barycentric orbital period, corrected for
the effect of solar radiation pressure on the disintegrated comet's headless
condensation, was \mbox{$683 \!-\! 41 = 642 \pm 9$ yr} and the time of the
previous perihelion passage was \mbox{AD~$2011 \!-\! 642 =$ AD~1369} with
the same uncertainty.  England (2002) lists a possible Kreutz sungrazer
seen in Korea and Japan in the western skies in early March 1368, Lovejoy's
potential parent.

An overwhelming majority of historical Kreutz sungrazers move in orbits that are
unknown, their sungrazing nature is merely suspected, and the populations they
belong to are indeterminate.  Only for a few of them did past investigations
show preference for Population~I or II.  It is obvious that preference for
Population~I over II also suggests preference for Population~Pe or Pre-I
over II, because one cannot distinguish between Populations~I, Pe, or Pre-I
among these comets.  Similarly, preference for Population~II over I may
actually imply a sungrazer of Population~IIa, etc.

In any case, the fundamental and practically the only parametric constraint
on the potential historical Kreutz sungrazers is time --- meaning their
(approximate) perihelion time --- as they become prominent, naked-eye
objects in the course of days or weeks after perihelion.  The distribution
of perihelion times of a generation of fragments is determined not only
by their orbital periods but also by the relations among the past perihelion
times and orbital periods of their parent sungrazers.  A representative
pedigree chart of the Kreutz system is assembled by consecutive generations
of fragments, whose birth dates are essentially given by their parents'
perihelion times.  Indeed, as nontidal fragmentation is known to have
a minor effect on the orbital period, the temporal dimension of the
pedigree chart is determined nearly exclusively by tidally-driven
fragmentation events.

Tidal disruption of a sungrazer was addressed in Papers~2 and 4 in the
case when its two fragments, A and B, acquire no differential momenta,
ending up in orbits of different periods, $P_{\rm A}$ and $P_{\rm B}$,
only because of a slight shift in the radial distance, $\Delta U(r_{\rm
frg})$, between their centers of mass after separation,
\begin{equation}
P_{\rm B} = P_{\rm A} \! \left[ 1 - \frac{2 \Delta U(r_{\rm
  frg})}{r_{\rm frg}^2} P_{\rm A}^{\frac{2}{3}} \right]^{-\frac{3}{2}}
  \!\!\!,
\end{equation}
where $r_{\rm frg}$ is the heliocentric distance at fragmentation.
In this equation both $\Delta U$ and $r_{\rm frg}$ are in AU, while
$P_{\rm A}$ and $P_{\rm B}$ are in years.  The difference
\mbox{$\Delta P_{\rm frg} = P_{\rm B} \!-\! P_{\rm A}$} is
approximated by
\begin{equation}
\Delta P_{\rm frg} = \frac{3}{r_{\rm frg}^2}
 P_{\rm A}^{\frac{5}{3}}\,\Delta U(r_{\rm frg}).
\end{equation}
Investigations of Roche's limit (e.g., Aggarwal \& Oberbeck 1974)
suggest that a body of tensile strength $T$ and diameter $\cal D$
begins to fragment tidally at a distance, $r_{\rm frg}$, from the
Sun, at which
\begin{equation}
{\cal D} = a_0 \sqrt{T} \, r_{\rm frg}^{\frac{3}{2}}
\end{equation}
or
\begin{equation}
{\cal D} = \frac{a \sqrt{T} \,r_{\rm frg}^{\frac{3}{2}}}{\sqrt{1 \!-\!
  b r_{\rm frg}^3}} \simeq a \sqrt{T} r_{\rm frg}^{\frac{3}{2}}
  \! \left( 1 \!+\! {\textstyle \frac{1}{2}} b r_{\rm frg}^3 \right),
\end{equation}
$a_0$, $a$, and $b$ are constants.{\vspace{-0.08cm}}  Next,
\mbox{$\Delta U(r_{\rm frg}) \sim {\cal D} \sim r_{\rm frg}^z$}, where
\mbox{$z \geq {\textstyle \frac{3}{2}}$}.  From Equation~(5) I get for
a given $P_{\rm A}$
\begin{equation}
\Delta P_{\rm frg} \sim \frac{\Delta U(r_{\rm frg})}{r_{\rm frg}^2}
  \sim r_{\rm frg}^{-y},
\end{equation}
where \mbox{$y = 2 \!-\! z \leq {\textstyle \frac{1}{2}}$} or near zero.
The implication is that $\Delta P_{\rm frg}$ is at worst only weakly
dependent on the heliocentric distance at fragmentation.

This conclusion might explain the tendency to a peculiarly regular interval
near 100~yr, with which orbital periods of neighboring fragments of a
tidally-disrupted sungrazer appear to differ.  The effect is perceived
among the four nuclear fragments of the Great September Comet of 1882,
whose osculating orbital periods equaled, respectively, 671.3$\,\pm\,$4.4~yr,
771.8$\,\pm\,$1.9~yr, 875.0$\,\pm\,$3.0~yr, and 955.2$\,\pm\,$6.0~yr
(Kreutz 1891), thus implying $\Delta P_{\rm frg}$ of 100.5$\,\pm\,$4.8~yr,
103.2$\,\pm\,$3.6~yr, and 80.2$\,\pm\,$6.7~yr, respectively.  In the
pedigree chart the increments in the orbital period are to be reckoned
from the parent object, $\Delta P_{{\rm p}\rightarrow{\rm f}}$. As the
period of the presplit comet was \mbox{$P_{\rm par} = 744.1$ yr}, the
values of $\Delta P_{{\rm p}\rightarrow{\rm f}}$ were $-$72.8~yr for the
first fragment, +27.7~yr for the second fragment, +130.9~yr for the third
fragment, and +211.1~yr for the last fragment.

Another data point of this kind has been provided by the pair of the 1882
comet and Ikeya-Seki as fragments of their common parent (Marsden 1967),
which implies \mbox{$\Delta P_{\rm frg} = 83.1$ yr}.  The resulting average
$\Delta P_{\rm frg}$ of $\sim$92~yr, when inserted with the relevant values
of \mbox{$P_{\rm par} \simeq 770$ yr} and \mbox{$r_{\rm frg} \simeq 0.0078$
AU} into Equation~(5), gives an average radial center-of-mass shift of
\mbox{$\Delta U \simeq 4.3$ km}.
%
% Values of $\Delta P_{\rm frg}$ from a slightly expanded range of
% $\sim$70~yr through $\sim$110~yr will be used as a selection
% criterion to fit into an extended pedigree chart of the Kreutz
% system as many suspected historical sungrazers as other constraints allow.

The objective of this discussion has been to call attention to possible
constraints that should be exploited in the process of setting up the
pedigree charts.  Application of these constraints is worked out in
Section~4.

\section{Cumulative Temporal Distribution of the Kreutz/Super-Kreutz
 System}
Before assembling examples of a pedigree chart, I maintain that its
presentation is aided by information on the cumulative temporal
distributions of potential historical Kreutz/super-Kreutz sungrazers
and historical comets in general.

In the following I use Hasegawa's (1980) catalogue of more than
1000~ancient and naked-eye comets to plot their cumulative distribution
(up to AD~1702), while for the historical Kreutz/super-Kreutz sungrazers
I employ two independent datasets:\ a list of early potential sungrazing
comets by England (2002), which contains 62~entries; and, separately, a
more restricted group of 24~possible sungrazers, for which Hasegawa \&
Nakano (2001) estimated the perihelion dates.

\begin{figure}
\vspace{0.2cm}
\hspace{-0.2cm}
\centerline{
\scalebox{0.62}{
\includegraphics{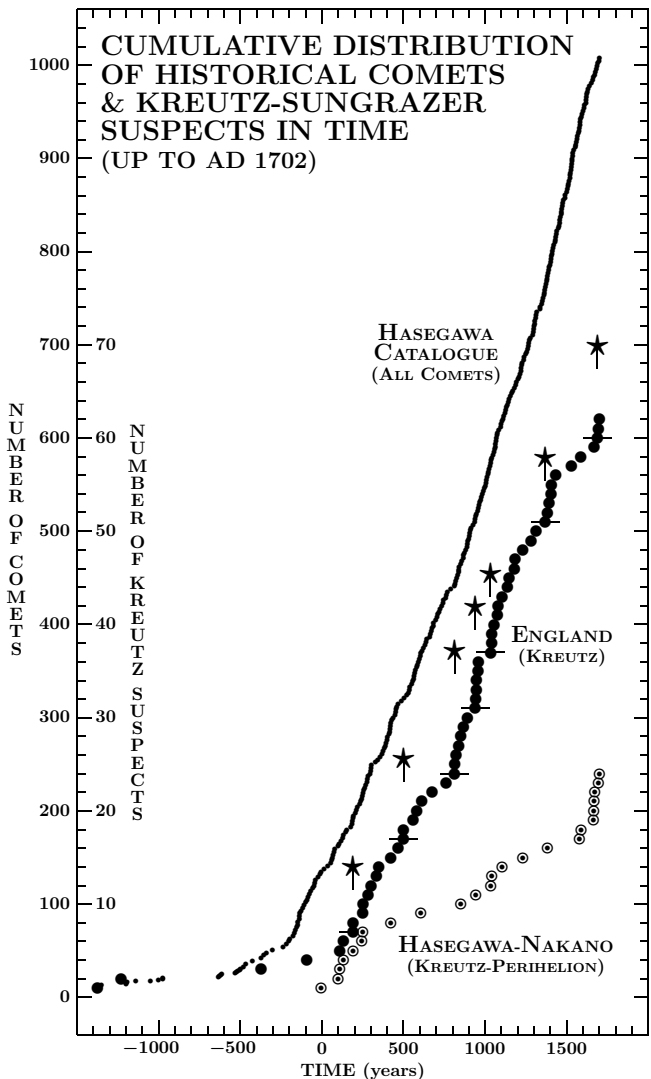}}} %  FIGURE 1
\vspace{-0.1cm}
\caption{Cumulative temporal distributions of historical comets from
 Hasegawa's (1980) catalogue (dots) and potential historical members
 of the Kreutz/super-Kreutz system from England's (2002) list (solid
 circles) and Hasegawa \& Nakano's (2001) list (circled dots).  Note
 that the scales for the sungrazers and the comets in general differ
 by a factor of 10.  The stars with the vertical and horizontal
 dashes mark the times of steep increase (bumps) on the England
 distribution curve.{\vspace{0.7cm}}}
\end{figure}

The three distributions of the historical comets are plotted in Figure~1. 
Even though the Hasegawa-Nakano collection is too small for statistical
purposes, it does support the more extensive set by England in showing
discrete points of sudden steep increase, referred to below as the
``bumps.''  The presence of these features, marked by the stars in
Figure~1, discriminates the distribution curve for the Kreutz/super-Kreutz
sungrazers from that for all comets, as no prominent bumps are apparent on
the latter curve.

From what was stated in Paper~1, the bumps are undoubtedly signatures
of fragmentation events at heliocentric distances larger than, say,
\mbox{5--10 AU} along the outgoing leg of the orbit.  Their sequence
on the distribution curve does not appear entirely random:\ of the six
intervals the bumps make up, the lengths of four are between 310~yr and
334~yr, or about four orbital-period differences $\Delta P_{\rm frg}$
discussed in Section~2.  The remaining two bump intervals equal,
respectively, 126~yr and 95~yr, the latter being remarkably close to
an average value of $\Delta P_{\rm frg}$.

The other major distribution signature is the general trend of the curves,
apparent from Figure~1:\ disregarding the bumps, the basic linearity of
the Kreutz/super-Kreutz curve, confirmed by a 60~percent error of the
coefficient of the tested quadratic term, contrasts with the clearly
perceived gradually increasing steepness of the curve of comets from the
Hasegawa catalogue.  Starting with the fifth entry of the England (2002)
dataset, the cumulative number, ${\cal N}_{\rm K}$, of the potential
historical Kreutz/super-Kreutz sungrazers is, as a function of time,
$t$, approximated by a least-squares fit as follows:
\begin{eqnarray}
{\cal N}_{\rm K}(t) & = & -1.04 + 3.763 \, t, \\[-0.1cm]
                 &   & \pm 0.69 \, \pm \! 0.067 \nonumber
\end{eqnarray}
where $t$ is expressed in centuries (e.g., AD~252 is given as \mbox{$t
= 2.52$}).  This equation can be rewritten as
\begin{equation}
{\cal N}_{\rm K}(t) = 3.763 \, (t - 0.276),
\end{equation}
which indicates that the date when \mbox{${\cal N}_{\rm K} = 0$}, or
the zero date of this cumulative distribution, equaled $t_{\rm K} =
0.276$, that is, August of AD~27.

In order to meaningfully compare the Kreutz/super-Kreutz cumulative
distribution with the cumulative distribution of all historical comets,
the serial numbers of the latter, $\cal N$, ought to be modified,
because Hasegawa's catalogue lists 135~comets at \mbox{$t < 0$}.  When
appropriately truncated, the cumulative distribution can be approximated
by the following cubic least-squares fit:
\begin{eqnarray}
{\cal N}(t) & = & -13.2 + 40.03 \, t - 0.49 \, t^2 + 0.0757 \, t^3.\\[-0.1cm]
	    &   &  \;\:\pm 1.9 \;\:\, \pm \! 0.91 \;\;\, \pm \! 0.12
	       \;\;\;\; \pm \! 0.0048 \nonumber
\end{eqnarray}
The solution shows that \mbox{${\cal N} = 0$} at \mbox{$t = t_0^\ast =
0.331$}, that is the zero date took place in February of AD~33, only
5.5~yr after the zero date of the adopted cumulative distribution of
the potential historic sungrazers.

I next proceed to rewrite Equation (11) in order that its form should
emulate an expanded version of expression (10), in particular,
\begin{equation}
{\cal N}(t) = \alpha (t \!-\! t_0) \! \left[ 1 + \beta (t \!-\! t_0) +
  \gamma (t \!-\! t_0)^2 \right].
\end{equation}
The condition is that \mbox{$t_0 = t_{\rm K}$}, while $\alpha$, $\beta$,
and $\gamma$ are constants that can be expressed via the coefficients in
Equation~(11), derived from the following considerations.

If a function $y(x)$ is given by a cubic polynomial, for \mbox{$y = 0$}
one deals with a cubic equation of the type
\begin{equation}
a_0 + a_1 (x \!-\! x_0) + a_2 (x \!-\! x_0)^2 + a_3 (x \!-\!x_0)^3 = 0,
\end{equation}
where $x$ is the variable, $x_0$ a constant, and $a_k\!$'s the
coefficients.  When \mbox{$x_0 = 0$}, a cubic equation becomes
\begin{equation}
b_0 + b_1 x + b_2 x^2 + b_3 x^3 = 0,
\end{equation}
where $b_k\!$'s are other coefficients.  Should the cubic equations
(13) and (14) be identical, the coefficients $b_k\!$'s are related to
$a_k\!$'s by{\vspace{-0.1cm}}
\begin{eqnarray}
b_0 & \:\!=\:\! & a_0 - a_1 x_0 + a_2 x_0^2 - a_3 x_0^3, \nonumber \\[-0.05cm]
	b_1 & = & a_1 - 2 a_2 x_0 + 3 a_3 x_0^2, \nonumber \\[-0.05cm]
	b_2 & = & a_2 - 3 a_3 x_0, \nonumber \\[-0.05cm]
b_3 & = & a_3.
\end{eqnarray}
Conversely, from Equations (15) the coefficients $a_k\!$'s are equal in
terms of $b_k\!$'s,{\vspace{-0.1cm}}
\begin{eqnarray}
a_0 & \:\!=\:\! & b_0 + b_1 x_0 + b_2 x_0^2 + b_3 x_0^3, \nonumber \\[-0.05cm]
	a_1 & = & b_1 + 2 b_2 x_0 + 3 b_3 x_0^2, \nonumber \\[-0.05cm]
	a_2 & = & b_2 + 3 b_3 x_0, \nonumber \\[-0.05cm]
a_3 & = & b_3.
\end{eqnarray}

It is apparent that to get the parameters $\alpha$, $\beta$, and $\gamma$
in Equation~(12) from the coefficients in Equation~(11) is equivalent to
getting $a_k\!$'s in (13) from $b_k\!$'s in (14) via the transformation
equations in (16).  Replacing ${\cal N}(t)$ with \mbox{${\cal N}(t) \!+\!
2.19$}, I find (12) in the form
\begin{equation}
{\cal N}(t) \!=\! 39.78 \:\! (t \!-\! t_0) \! \left[ 1 \!-\! 0.01074
  (t \!-\! t_0) \!+\! 0.00190 (t \!-\! t_0)^2 \right] \! ,
\end{equation}
where, as stipulated, \mbox{$t_0 = t_{\rm K} = 0.276$}.

With \mbox{$t = 17.02$}, Equations (10) and (17) give, respectively,
\mbox{${\cal N}_{\rm K} = 63$} and \mbox{${\cal N} = 901$} for the
period of time AD~27 through 1702, close to the expected totals.  The
ratio $\Re$ of the number of potential historical Kreutz/super-Kreutz
sungrazers in England's (2002) paper to the number of all historical
comets in Hasegawa's (1980) catalogue is thus 63/901, almost exactly
0.07.  It is interesting to compare this result with a similar fraction
for recent {\it definite\/} Kreutz sungrazers, $\Re_0$, because an
assumption that $\Re$ and $\Re_0$ should remain essentially constant
with time offers an estimate for the ratio $\Re_0/\Re$ of definite
sungrazers to potential Kreutz/super-Kreutz sungrazers, a number that
must be smaller than unity.

I comment on two examined datasets.  The first one, by Bortle (1998),
related to comets observed between 1800 and 2000, which at peak
brightness were near apparent visual magnitude 0 or brighter.  Among
40~comets in this category, Bortle had six Kreutz sungrazers, which
implied an unexpectedly high ratio of \mbox{$\Re_0 = 0.15$}.  The
result reflected the fact that objectively there is a~much higher
fraction of Kreutz sungrazers among comets brighter than magnitude~0
than among all naked-eye comets making up Hasegawa's catalogue.

The other dataset, by Green (2020), listed comets~be\-tween
1935 and 2020, which at peak brightness~were~of
apparent visual magnitude 3.9 or brighter.  Among the
48~tabulated comets were three sungrazers (Ikeya-Seki C/1965~S1,
White-Ortiz-Bolelli C/1970~K1,~and~\mbox{Lovejoy} C/2011~W3), giving
\mbox{$\Re_0 = 0.064$}.  However, since Pereyra C/1963~R1 (with a peak
magnitude of 2) was missed, it was in fact \mbox{$\Re_0 = 0.085$}.
When this dataset is truncated to include only comets with
apparent magnitude~0 at peak brightness, two sungrazers remain
among 13~comets, a ratio $\Re_0$ of 0.15, in full agreement with Bortle's
value!

The point in which these two sets differ dramatically from the sets of
historical comets is the contribution from southern-hemisphere observers,
which was essentially absent before the 19th century.  This point is
important because most Kreutz sungrazers are observable only from the
southern hemisphere.  As a result, the values derived for $\Re_0$ of
recent Kreutz sungrazers are much higher than in historical times.
The magnitude of this effect is very hard to estimate.  The problem is
further aggravated by a low number of Kreutz comets in the datasets.

In an effort to overcome, or at least mitigate, these difficulties and
augment the statistics as much as reasonably possible, I expand the
time interval~to~\mbox{1800--2020}, by combining the time intervals
from Green's as well as Bortle's datasets.  Of the nine naked-eye Kreutz
sungrazers seen during the 220~years (including the Eclipse comet
X/1882~K1 in addition to the eight obvious ones), I estimate that
only four brilliant daylight and/or twilight objects --- C/1843~D1,
C/1882~R1,~C/1965~S1, and C/2011~W3 --- could be discovered from
either~hemisphere.  Given the total number of comets in Hasegawa's
catalogue between AD~27 and 1702, the equivalent~num\-ber in 220~years
would be 118, and therefore the ratio \mbox{$\Re_0 = 4/118 \simeq 0.034$}.
As \mbox{$\Re = 0.07$}, the ratio \mbox{$\Re_0/\Re \simeq 0.5$}:\ {\it every
second object among England's 62~potential Kreutz sungrazers should
{\vspace{-0.05cm}}actually be a member of the system\/}.  I note that
{\vspace{-0.09cm}an error of $\pm$1~sungrazer implies errors of
$_{-0.009}^{+0.008}$ in $\Re_0$
and $_{-0.13}^{+0.11}$ in \mbox{$\Re_0/\Re$}.

Interestingly, when the number of all comets prior to 1702, discovered
from the northern hemisphere and catalogued by Hasegawa, is converted
to Green's (2020) interval of 85~years, it implies a total number of
46~comets. Green's list contains 48~objects down to peak magnitude
3.9, discovered by observers in both hemispheres.  It therefore
appears that Hasegawa's catalogue may include comets with peak
brightness as faint as apparent magnitude $\sim$4.5, if the rate
of arriving comets does not vary significantly with time.

The cumulative distribution of comet discoveries based on Hasegawa's
(1980) catalogue, leaves the issue of rate variability open.  In any
case, the catalogue-based distribution, whose slope increases with time,
contrasts in Figure~1 with the linearity of the cumulative distribution
of the potential historical Kreutz/super-Kreutz sugrazers in England's
(2002) study.  I find it highly unlikely that the progressively growing
rate of discovery in Hasegawa's catalogue could be attributed to any
effect other than gradually improved records on comet discoveries in
more recent times and an obviously higher rate of record survival than
that of older documents.  The question is why the distribution of
the Kreutz/super-Kreutz sungrazers varies with time linearly, failing
to show a similar effect?\,\,\,

A rather bold hypothesis that I offer in this context is based on
the assumption that the {\it intrinsic\/} cumulative distribution
of potential historical Kreutz/super-Kreutz sungrazers was in fact
approximated by a similar cubic polynomial, but subsequently was
gradually altered by the process of fragmentation, especially in close
proximity of perihelion, resulting over many centuries in the {\it
modified\/} linear distribution, as determined from Figure~1.  Fragmentation
was surely {\it augmenting\/} the number of Kreutz/super-Kreutz sungrazers
on the one hand (such as in the cases of C/1882~R1 and C/1965~S1), but
simultaneously decreasing their average dimensions on the other hand.
The number of disintegrating objects (such as C/1887~B1 or C/2011~W3)
was thus increasing with time, {\it lowering\/} the overall number of
surviving sungrazers.  Accordingly, measuring the modifications of
the distribution from a starting point at time \mbox{$t \rightarrow
t_{\rm K}$}, I stipulate that the intrinsic cumulative distribution
of potential historical Kreutz/super-Kreutz sungrazers in England's
dataset be approximated by a cubic polynomial of type (12), which is
a hybrid of the expressions (10) and (17),
\begin{equation}
{\cal N}_{\rm K}^{\!\:int}\:\!\!\!=\!3.763(t\!-\!t_{\rm K})\!\!\!\:\left[1\!-\!
 0.01074 (t \!-\! t_{\rm K})\!+\!0.00190 (t\!-\!t_{\rm K})^2\right]\!\!\!\:.
\end{equation}
The respective modified cumulative distribution, $\cal N_{\rm K}^{\it
mod}\!$, is given} by (10).  The difference,
\begin{eqnarray}
\Delta {\cal N}_{\rm K}(t) & = & {\cal N}_{\rm K}^{\it mod} \!-\!
  {\cal N}_{\rm K}^{\it int} \nonumber \\
		& = & 0.0404 \, (t \!-\! t_{\rm K})^2 \!\!\: \left[ 1 \!\!\: -
  0.177 \, (t \!-\! t_{\rm K}) \right],
\end{eqnarray}
describes the presumed effect of the fragmentation process on the
cumulative distribution of the Kreutz/super-Kreutz system in England's
dataset.  Since at AD~1702 (for \mbox{$t \simeq 17$}) Equation~(19) shows
that \mbox{$\Delta {\cal N}_{\rm K} \simeq -22$}, the disintegration
effect of the fragmentation process even\-tu\-ally did, after centuries,
prevail and the number of surviving sungrazers continued to drop with
time.  However, \mbox{$\Delta {\cal N}_{\rm K} > 0$} until about AD~590,
even though very slightly (Figure~2).  In this particular scenario, the
initial ratio \mbox{$\Re = 0.095$} and only about 36~percent of the
suspected historical Kreutz/super-Kreutz comets in England's dataset
would actually be members of the system.
\begin{figure}[t]
\vspace{0.15cm}
\hspace{-0.15cm}
\centerline{
\scalebox{0.73}{
\includegraphics{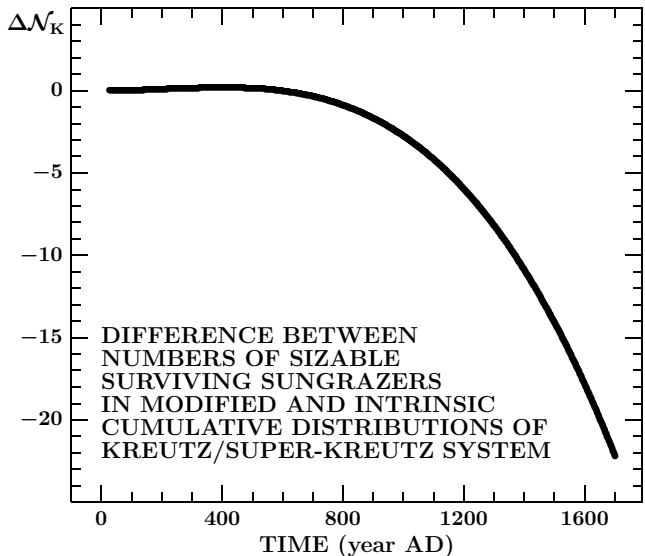}}}
\vspace{-0.1cm}
\caption{Hypothetical difference $\Delta {\cal N}_{\rm K}(t)$ between
the numbers of surviving sizable sungrazers from AD~27 until AD~1702, if
the intrinsic cumulative distribution follows that of other historical
comets in Hasegawa's (1980) catalogue and is modified by the process
of fragmentation in close proximity of perihelion to vary linearly
with time.  Starting with a set of 84~potential Kreutz/super-Kreutz
sungrazers between AD~27 and 1702 (intrinsic distribution), the model
predicts that 22~objects would disintegrate to leave a total of
62~surviving objects in England's (2002) dataset.{\vspace{0.6cm}}}
\end{figure}

\section{Representative Extended Pedigree Chart\\for the Kreutz System}
% Section 4
Accepting that 50~percent, or possibly more, of potential historical
Kreutz/super-Kreutz sungrazers are in fact unlikely to be members of
the system, I am now ready to assemble a representative extended pedigree
chart, first for the Kreutz system proper.

I call the pedigree chart {\it representative\/}, because with the
exception of the several definite members I have no means to make sure
that any selected potential sungrazer is the correct one --- it only
satisfies the stipulated conditions.  For several sungrazers,
whose status is superior to {\it potential\/} but inferior to {\it
definite\/} --- such as C/1668~E1, C/1689~X1, C/1695~U1, X/1702~D1,
X/1882~K1, and, of course, X/1106~C1 and the Chinese comet of 1138 ---
the population membership is believed to be known.

I call the pedigree chart {\it extended\/}, because it reaches to
the very origin of the Kreutz system, with a maximum possible number
of potential historical Kreutz sungrazers that satisfy the stipulated
conditions.  These objects are taken primarily from the papers
by Hasegawa \& Nakano (2001) and by England (2002), but also
included are ``sun-comets'' by Strom (2002) and additional objects,
such as the Eclipse Comet X/1882~K1 (Abney \& Schuster 1884; Marsden
1989).  The last category is that of ``missed'' sungrazers; they are
used in the chart~as~necessary to fill the gaps, but very sparingly.

The only constraints employed in assembling the pedigree chart derive
from the approximate conditions discussed in Section~2.  First of all,
limited evidence shows a tendency for the orbital periods of major
neigboring, tidally-generated fragments of their common parent sungrazer to
differ, very crudely, by \mbox{$\Delta P_{\rm frg} \simeq 100$ yr}.  Second,
judging from the near-perihelion fragmentation of the Great September Comet
of 1882, the range of the fragments' orbital periods, $P_{\rm frg}$, from
Kreutz's (1891) computations is equal to \mbox{$P_{\rm par} - \Delta
P_{\rm frg} \leq P_{\rm frg} \leq P_{\rm par} + 2 \, \Delta P_{\rm frg}$}.
Third, estimating that the surviving, massive sungrazers (with a potential
to break up into large enough fragments at their next return to perihelion)
have their orbital~periods typically in a range \mbox{$600 \; {\rm yr} <
P_{\rm par} < 800$ yr},~I~find for fragments' orbital periods to be confined
to an overall range \mbox{$\sim \! 500 \; {\rm yr} < P_{\rm frg} < \sim \!
1000$ yr}.  And fourth, the final constraint follows from an obvious
condition that the mass-weighted distribution of the orbital-period
differences, $\Delta P_{{\rm p} \rightarrow {\rm f}}$, relative to the
parent's orbital period --- a function of the fragments' center-of-mass
distribution --- should be centered on zero and the most massive fragments
(such as X/1106~C1 among the third-generation objects or C/1843~D1 among
the fourth-generation objects) should have their values of $\Delta P_{{\rm
p} \rightarrow {\rm f}}$ close to zero.  From the above constraints on
$P_{\rm par}$ and $P_{\rm frg}$ the overall condition on $\Delta P_{{\rm
p}\rightarrow{\rm f}}$ becomes
\begin{equation}
-300 \; {\rm yr} < \Delta P_{{\rm p}\rightarrow {\rm f}} < +400 \; {\rm yr}.
\end{equation}
At \mbox{$r_{\rm frg} = 0.0078$ AU}, this range of $\Delta P_{{\rm p}
\rightarrow {\rm f}}$ corresponds to a range \mbox{$-14 \; {\rm km} <
\Delta U < +19 \;$km}, or a total of 33~km, a shift whose magnitude is
well within a nucleus of 50~km across.  On the one hand, it should
be recalled that this number varies as $r_{\rm frg}^2$;\,on the other
{\vspace{-0.03cm}}hand,\,this momentum-free effect on an orbital period
is a theoretical construct that is merely a convenient approximation to
genuine conditions in the course of a tidal-fragmentation event, as in
fact each fragment acquires a separation velocity, however low it may be.

Except for the proper compartmentalization of the definite members of
the Kreutz system and the described constraints, no other conditions are
employed.  The list~of all entries in the pedigree chart for the Kreutz
system is presented in Table~1 and the chart itself is in Table~2.

\begin{table*}
\vspace{0.15cm}
\hspace{-0.18cm}
\centerline{
\scalebox{0.982}{
\includegraphics{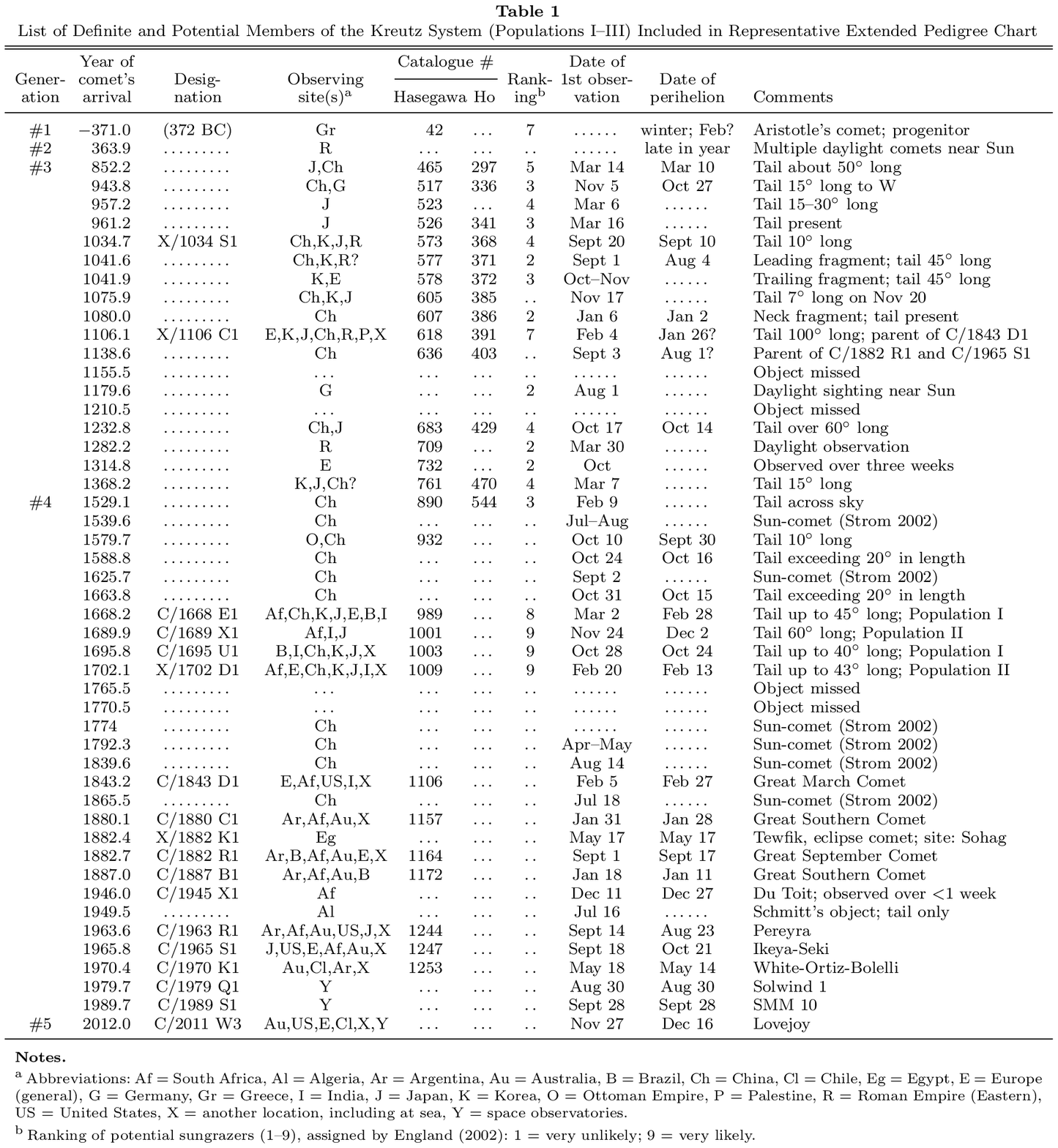}}} %  Table 1  (list.eps)
\vspace{0.75cm}
\end{table*}

Both tables are organized by a generation rank, starting with the
progenitor of the contact-binary model --- Aristotle's comet of
372~BC, which occupies the first row of Table~1.  Its successive
columns then offer the year of appearance; the designation (when
available); the country or countries (or otherwise determined
locations) from which the object was observed; the object's
serial numbers in, respectively, Hasegawa's (1980) and Ho's (1962)
catalogues; England's (2002) ranking; the date of the first
observation; the (often estimated) date of perihelion passage;
and the comments.

In Table 2, the perihelion time of the progenitor, the sole first-generation
object, is the only entry in column~1.  The second-generation objects
occupy columns~2--4:\ the orbital period, $P$, equaling the difference
between the perihelion times of the objects of the second and first
generations (column~3 minus column~1), in column~2; the perihelion time of
each second-generation object in column~3; and the letter designation of
this object as a fragment of the first-generation object, in column~4.~The
third-generation objects occupy four columns, 5--8:\ the orbital
period, $P$, equaling the difference between the perihelion times of the
objects of the third and second generations (column~7 minus column~3),
in column~5; the difference, $\Delta P$ (abbreviated from $\Delta P_{{\rm
p} \rightarrow {\rm f}}$ in Section~2), between the orbital periods
of the objects of the third and second generations (column~5 minus
column~2) in column~6; the perihelion time of each third-generation
object in column~7; and the letter designation of this object as
a fragment of the second-generation object, in column~8.  The
fourth-generation objects also occupy four columns, 9--12, which
are equivalent to columns~5--8 for the third-generation objects.
The last column~13 offers information on the structure of the Kreutz
system.

\begin{table*}
\vspace{0.1cm}
\hspace{-0.2cm}
\centerline{
\scalebox{1}{
\includegraphics{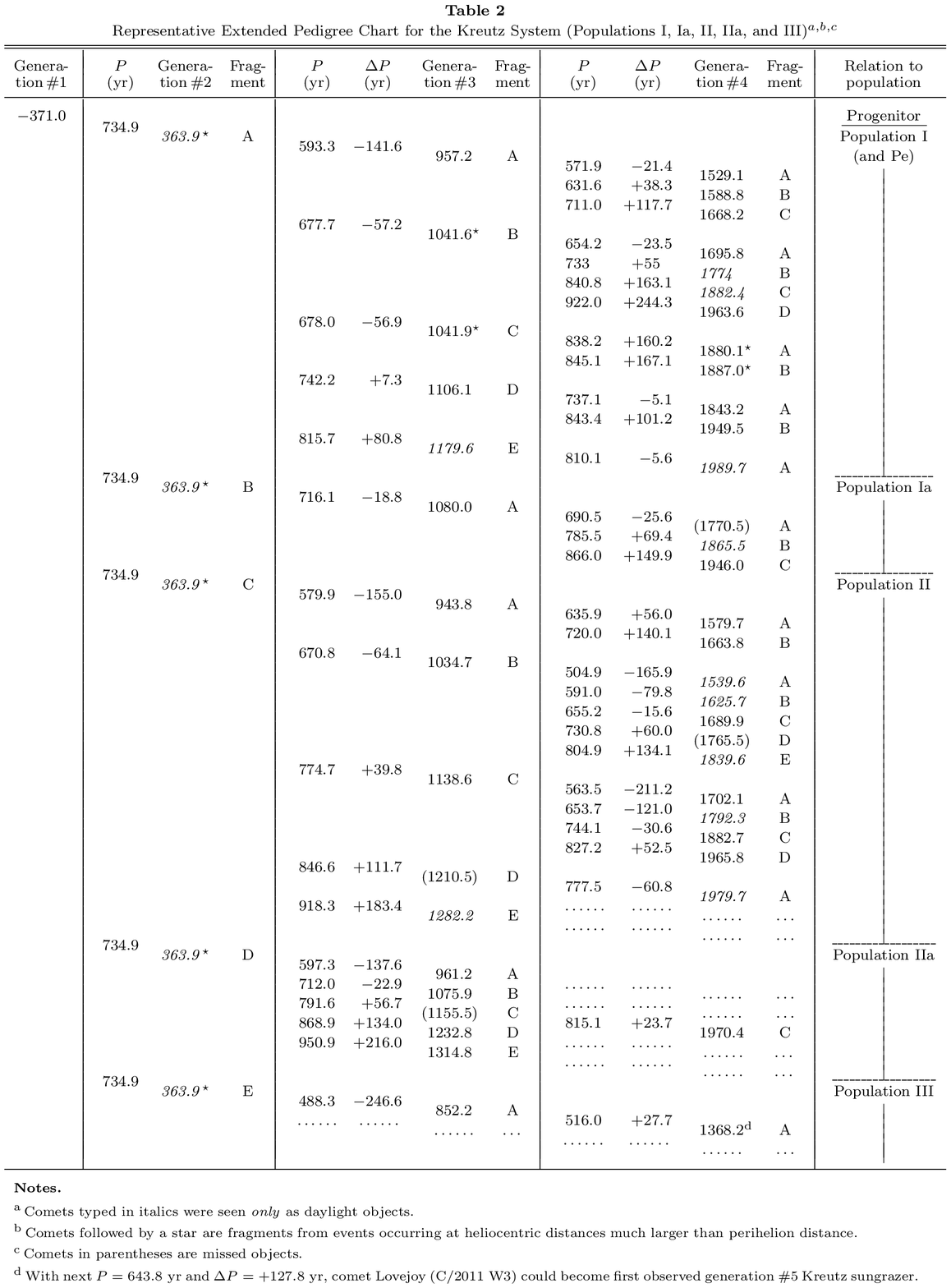}}} %  Table 2  (tx_363.eps)
\end{table*}

Based on the populations detected in Paper~1 among the SOHO Kreutz
sungrazers, a scenario of near-aphelion fragmentation of the progenitor's
two lobes and neck into ten major fragments, arriving within days of one
another at perihelion in AD~363 as the second-generation objects, was
simulated in Paper~3.  Of this cluster of arriving sungrazers only a
fraction is listed in the pedigree chart in Table~2:\ Fragment~A is
a precursor of Populations~I and Pe (from Lobe~I), Fragment~B of
Population~Ia (from the neck), Fragment~C of Population~II, Fragment~D
of Population~IIa, and Fragment~E of Population~III (all three from
Lobe~II).

Each of these second-generation objects (or the first-generation
fragments) is next presumed in the pedigree chart in Table~2 to have
fragmented into third-generation objects (or second-generation
fragments).  The arriving Fragment~A split into four major pieces,
but one of them broke up again into two later, far from perihelion.
Accordingly, a total of five major fragments returned to perihelion
between the mid-10th and late 12th centuries.  First was the comet
of 957 (a second-generation Fragment A), two comets of 1041, one in
September, the other in November (second-generation Fragments~B and C,
respectively), the Great Comet of 1106 or X/1106~C1 (a second-generation
Fragment~D; the most massive subfragment of the arriving Fragment~A
and the parent to the Great March Comet of 1843), and the daylight
comet of 1179 (a second-generation Fragment~E).

The dimensions of the arriving Fragment~B, the presumed product of
the progenitor's neck in the putative cluster of sungrazers, should
have been relatively small.  On the assumption that it lost little mass
at perihelion, it should have returned to next perihelion with only a
slightly changed orbital period as the comet of 1080.

On the other hand, the arriving Fragment C is presumed to have split
into five sizable subfragments at perihelion in AD~363, which returned
between the mid-10th and late 13th centuries.   First to come back was
the comet of 943 (another second-generation Fragment~A).  Next was the
comet of 1034 (another second-generation Fragment~B), then the most
massive of this group, the comet of 1138 (another second-generation
Fragment~C; the parent of the Great September Comet of 1882 and
Ikeya-Seki), followed by a comet of 1210 (another Fragment~D; missed)
and by the comet of 1282 (another second-generation Fragment~E).

Next I assume that when the first-generation Fragment D arrived at
perihelion in AD 363, it also broke up into five second-generation
fragments.  In the order of increasing orbital period they were the
comets of 961 (Fragment~A), 1075 (Fragment~B), 1155 (Fragment~C;
missed), 1232 (Fragment~D), and 1314 (Fragment~E).  Strangely, the
missed sungrazer of 1155 is a candidate for the parent to comet
White-Ortiz-Bolelli (C/1970~K1).

The pedigree chart in Table 2 does not offer any detailed fragmentation
history of the remaining sungrazers from the cluster arriving in AD~363,
namely, Fragments~E, F, \ldots (precursors to Populations~III, IIIa,
\ldots).  For the first-generation Fragment~E, the single presumed
second-generation fragment is the comet of 852, as it may have been,
via the comet of 1368, a presursor of comet Lovejoy (C/2011~W3), the
only fifth-generation object under consideration.

The subsequent part of the pedigree chart in Table 2 presents fragmentation
products of the arriving second-generation fragments (Generation \#3).
For example, for Population~I (and Pe), the comet of 957 --- with the
shortest orbital period --- is assumed to have split at perihelion into
three third-generation fragments, giving birth to the comets of 1529
(Fragment~A; listed by England); 1588 (Fragment~B; listed by Hasegawa
\& Nakano), and 1668 (Fragment~C; well-known as a potential Kreutz
sungrazer C/1668~E1).

The September comet of 1041 is proposed to have been the parent to four
sungrazers:\ the comet of 1695 (also known as C/1695~U1; Fragment~A),
the sun-comet of 1774 (Fragment~B), the Eclipse Comet of 1882 (Fragment~C),
and comet Pereyra C/1963~R1 (Fragment~D).  The strongly asymmetric
distribution of $\Delta P$ for the products of this 1041 comet suggests
that additional unrecorded fragments should have arrived in the early
17th century and/or late 16th century.

The November comet of 1041 is proposed to have broken up far from
perihelion to generate the pair of bright but not very massive 19th
century sungrazers C/1880~C1 (Fragment~A) and C/1887~B1 (Fragment~B),
in compliance with the scenario presented in Paper~1.  However, to
make it plausible requires, in the least, one major sungrazer that
arrived at perihelion around 1720 but was apparently missed.  In any
case, the fragmentation history of this second 1041 comet is very much
incomplete in Table~2.

The famous comet of 1106 (X/1106 C1) did not necessarily produce any
major fragments, returning as the Great March Comet of 1843, as already
noted.  A much less prominent fragment could have been the possible
Kreutz member reported by Schmitt (1949).  The parent did however
fragment profusely, producing much of the current stream of dwarf
sungrazers, observed by the SOHO spacecraft.

One of the sungrazers discovered in the images taken by the Solar
Maximum Mission, SMM~10, may be a fragment of the daylight comet
of 1179.  There could have been other fragments in the past, in
particular around 1900, which were missed.  This daylight comet
is one of those that are expected to potentially produce bright
sungrazers in the course of the 21st century.

There is no need to continue listing the birth of the third-generation
fragments (Generation \#4) from the second-generation fragments in
Populations~Ia, II, IIa, and III, as the process is similar to that
in Population~I.  The implications of the pedigree chart are presented
next.

\section{Implications of the Pedigree Chart\\for the Kreutz System}
% Section 5
The most obvious outcome of the process of tidal~fragmentation is
rapid diffusion of Kreutz sungrazers in time.  Table~1 shows a
minimum perihelion-time separation of 735~yr between Generations~\#1
and \#2, 488~yr between Generations~\#2 and \#3, 161~yr between
Generations~\#3 and \#4, and merely 22~yr between Generations~\#4
and \#5.  This diffusion effect is so overwhelming that several
revolutions after their birth, Kreutz sungrazers should have an
essentially random distribution of their perihelion times.  Given
that a tendency to clustering in time~is still apparent among the
observed objects, the first conclusion is that the {\it Kreutz
system has to be of a very~young age\/}, a condition that the
contact-binary model~is~entirely consistent with.

\begin{figure}
\vspace{0.16cm}
\hspace{-0.16cm}
\centerline{
\scalebox{0.723}{
\includegraphics{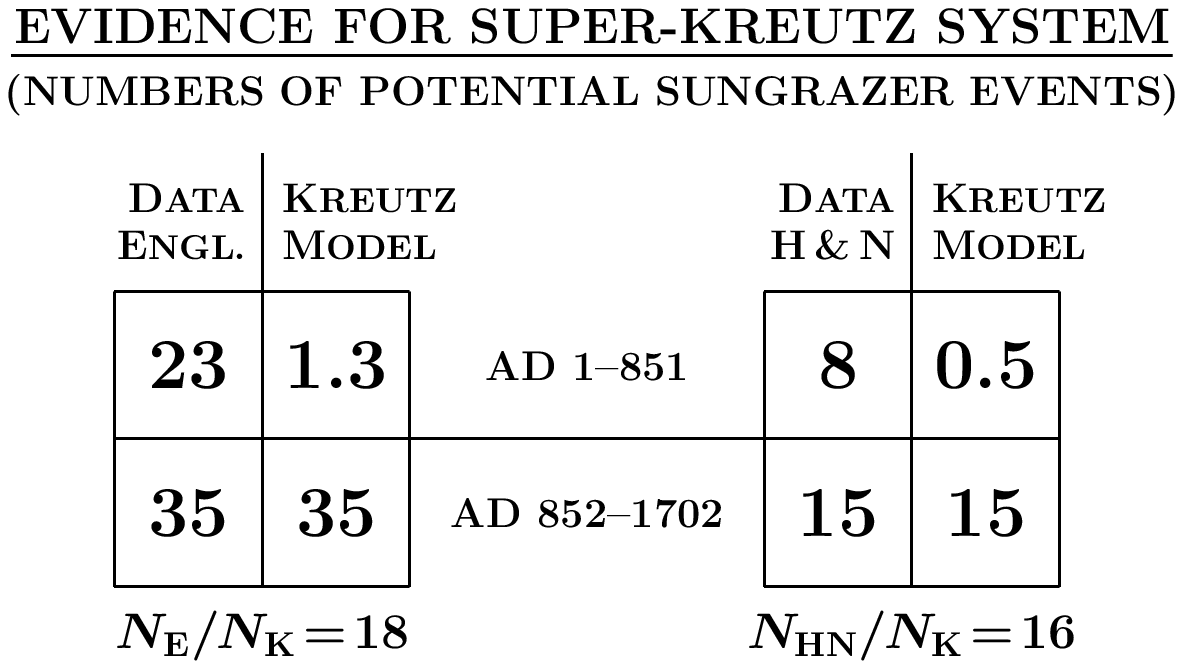}}} %  Figure 3
\vspace{-0.05cm}
\caption{Comparison of the numbers of potential historical Kreutz
sungrazers in England's set, $N_{\rm E}$ (left), and Hasegawa \&
Nakano's set, $N_{\rm HN}$ (right), with the normalized numbers of
objects from the pedigree chart of the Kreutz system, $N_{\rm K}$.
Note the considerable excess number of comets in either set over
the normalized number from the pedigree chart in the period of
\mbox{AD~1--851} relative to that in the period of AD~852--1702
(\mbox{$N_{\rm E}/N_{\rm K} = 23/1.3 = 18$}, \mbox{$N_{\rm
HN}/N_{\rm K} = 8/0.5 = 16$}), presenting strong evidence for
the existence of the super-Kreutz system.{\vspace{0.8cm}}}
\end{figure}

To continue along these lines, one can divide an appropriate time
interval, over which the several generations of potential Kreutz
sungrazers are shown in the pedigree chart to have been coming to
perihelion, into equal parts and compare the relative numbers, as
organized in Table~1, with the numbers of the potential historical
Kreutz sungrazers in England's (2002) and Hasegawa \& Nakano's (2001)
sets.  For example, adopting an overall period of time AD~1 through
AD~1702 (which happens to be the end year of either set) and dividing
it into two half-periods of AD~1 through AD~851 and AD~852 through
AD~1702, Table~1 shows that there was a single sungrazer event
in the first half-period, but 28 events (852, 943, \ldots, 1695,
1702) in the second.  In relative terms, these numbers represent
3.4~percent and 96.6~percent, respectively.

Comparison of their normalized numbers with the respective numbers of
potential historical Kreutz sungrazers in England's and Hasegawa \&
Nakano's sets is offered in Figure~3.  Immediately obvious is an
enormous excess of objects in either of the two datasets relative
to the number required by the pedigree chart of the Kreutz system
in the early period (\mbox{AD 1--851}), as opposed to the later
period (\mbox{AD 852--1702}).  Either dataset is found to contain a
greater number of potential Kreutz sungrazers than indicated by
the pedigree chart by more than a factor of ten!

Given that in the contact-binary model the single relevant sungrazer
event in the period of \mbox{AD 1--851} is that of the Ammianus
daylight comets, not listed by either England or Hasegawa \& Nakano,
the situation is in fact even more extreme than portrayed in Figure~3:\
{\it none\/} of the potential sungrazers before AD~852 presented in
either dataset is a {\it member of the Kreutz system proper\/}.
This fact represents remarkably robust evidence of the {\it existence
of the super-Kreutz system\/}:\ its members include all genuine
sungrazers among the 23~objects in England's dataset and among the
eight objects in Hasegawa \& Nakano's dataset.

In reviewing the condition for $\Delta P$ (i.e., $\Delta P_{{\rm p}
\rightarrow {\rm f}}$), described in Section~4, Table~2 shows that
in the majority of cases the value of
\begin{equation}
\Delta P_{\rm frg} = \Delta P_{{\rm p} \rightarrow {\rm f}_{k+1}}
 - \Delta P_{{\rm p} \rightarrow {\rm f}_k}
\end{equation}
for two neighboring fragments, f$_k$ and f$_{k+1}$, is in a general
range of 70~yr to 90~yr.  Two outstanding exceptions, the two comets
of 1041 and the pair of C/1880~C1 and C/1887~B1 concern fragments
that separated far from perihelion and are marked with stars in
Table~2.

There is some confusion about the comet(s) of 1041.  Both England
(2002) and Ho (1962) list two objects, one in September, the other
in November.  They also agree that either comet was observed in
Korea.\footnote{England calls the first comet Korean-Byzantine; Ho,
as a rule, lists only events recorded in Oriental sources.}  On~the~other
hand, Hasegawa~\&~Nakano~(2001) consider the appearance of only one
comet, which was observed from September through November.  They refer
to it as a Chinese-Korean object because they offer an additional
account from the Beijing Observatory, not available until 1988.
A problem with the assumption of a single object is the unusually
long period of observation, which in turn requires a considerable
intrinsic brightness (absolute magnitude of $-$1, by far the
brightest in Hasegawa \& Nakano's set).  The premise of a single comet
in late 1041 thus entails the need for its anomalous properties.
On the other hand, the scenario of two sungrazers about two months
apart at perihelion requires their breakup at a moderately large
heliocentric distance, and appears to be preferable.

Comet Lovejoy's (C/2011 W3) pedigree presents a major challenge.
While in Section 2 I suggested the comet of 1368 as a possible
parent, a fourth-generation object (or a third-generation fragment)
of the Kreutz system, its own pedigree is a puzzle.  To fit the
required temporal sequence, its barycentric orbital period would
have to be only marginally longer than 500~yr and its parent (a
third-generation object) would have to have a still shorter period
of less than 500~yr!  Either period is at the lower boundary of the
estimated range (Section 4) and therefore suspect.  If the \mbox{1368
$\Rightarrow$ 2011} link is valid, but the preperihelion period of
the 1368 comet is longer, comet Lovejoy would be a member of the
super-Kreutz system but not the Kreutz system proper (Section~6).

The attentive reader will notice that the scenario involving comet
Pereyra (C/1963~R1) described in this paper differs slightly from
the scenario proposed for this sungrazer in my previous papers (e.g.,
Paper~2).  The difference is in the history of its precursors, as
the pedigree chart in Table~2 assumes that Pereyra's direct parent
fragmented {\it after\/}, rather than {\it before\/}, AD~363,
a variation in the orbital evolution that is dynamically insignificant.
The only critical point was to find a path to separate a precursor of
comet Pereyra from precursors of the Great March Comet of 1843 on the
one hand and the Great Southern Comet of 1880 on the other hand. Either
scenario satisfies this condition.

In summary, the representative pedigree chart in Table~2 contains
20 of the 62 potential historical sungrazers in England's (2002) set,
12~of~the~24~such~objects~in Hasegawa \& Nakano's (2001) set,
and~6~of~the~14 sun-comets in
Strom's~(2002)~set.~The~chart~shows~four~suc\-cessive
generations of objects~and~hints~at~a~\mbox{single}~case of the fifth
generation object.~In~certain~\mbox{directions}~the fragmentation-driven
divarication is~described~to~considerable~extent,~such~as~in~the~cases~of~the~\mbox{third-generation}
Fragment~B~branch~of~\mbox{Population}~I~or~\mbox{Fragments}~B~and~C branches
of~Population~II.~In~other~cases,~such~as~in~the third-generation
Fragments~C,~D,~and~E~\mbox{branching}~of Population~I
or~\mbox{Fragments}~A~and~D~\mbox{branching}~of~Population~II,
the divarication~is~\mbox{outlined}~\mbox{incompletely}.~Only selected
\mbox{fourth-generation}~\mbox{objects}~are presented~in~the chart for
Populations~Ia, IIa,~and~III,~\mbox{whereas} the pedigree links in 
Populations \mbox{Pre-I}, IIIa, and IV are not fully pursued in Table~2.

My final comment~in~this~section~is~a~\mbox{reiteration}~that
fragmentation~paths in the~chart~are~arbitrary, except~for
valid~constraints~and~the~links~that~define~the~contact-
binary model, namely, $-$371 $\!\Rightarrow\!$ 363-A $\!\Rightarrow\!$
1106-D $\!\Rightarrow\!$ \mbox{1843-A} and \mbox{$-$371 $\!\Rightarrow\!$
363-C $\!\Rightarrow\!$ 1138-C $\!\Rightarrow\!$ 1882-C}.

\section{Representative Extended Pedigree Chart\\for a Super-Kreutz System}
% Section 6
The primary objective of this section is to show that, in investigating
the process of cascading fragmentation, a fair number of the potential
historical sungrazers (especially the objects from the first millennium
AD) listed in the datasets by England (2002), Hasegawa \& Nakano (2001),
Strom (2002), and others, which were not readily accommodated by the
pedigree chart for the Kreutz system proper in Table~2, could successfully
be fitted into a representative extended pedigree chart for a super-Kreutz
system.

I first describe the temporal range of the sungrazers under consideration
for this chart.  The orbits of the Great March Comet of 1843 and the Great
September Comet of 1882 were in Paper~4 integrated back to 5~BC, the time
of the contact-binary's bifurcation near aphelion.  Further integration
of the progenitor's orbit into the past then continued over another
2$\frac{1}{2}$ revolutions about~the Sun, determining in the process the
times of the consecutive perihelion passages in the years of 372~BC (or
$-$371), 1125~BC (or $-$1124), and 1901~BC~(or~$-$1900).  For inclusion
in the chart, the comet is assumed to have passed through perihelion in
these particular years even if there is no record about its possible
sighting.  There is of course nothing sacred about the predicted return
of $-$1900 and the orbit could have been integrated another revolution
or two about the Sun further back in time.  In practice, however, this would
make little sense because one has essentially no information available
on comets before the 2nd millennium BC.

Two issues associated with these early times of comet appearances were
briefly addressed in Paper~4 and are reiterated here.  One was an
uncertainty of dozens of years in dating most events, especially in
the 2nd millennium BC.  Granting this ambivalence, the comet that
allegedly was seen in Sumer around $-$1930 may have~been the sungrazer
described by Aristotle two revolutions~about the Sun later, in $-$371.
The respective entries in Hasegawa's (1980) catalogue of ancient comets
show Pingr\'e's (1783) cometography as the source of information.

The other issue involved a speculation about the Greek comet of $-$1200
or $-$1175 in Hasegawa's catalogue as a potential appearance of an early
fragment of the $-$1900 comet.  Here I extend the range of conjectures
of this kind to include:\ (i)~a Chinese comet that appeared according to
Ho (1962) in one of the years $-$1121, $-$1108, $-$1054 (most likely),
or $-$1029; (ii)~a possible comet during a total solar eclipse, etched
on an oracle bone discovered near Anyang, Henan Province, northern
China (England 2002), dated from $-$1225 or $-$1197 or $-$1162 or
$-$1160,\footnote{See website
{\tt https://solarsystem.nasa.gov/eclipses/about-}{\linebreak}{\tt
eclipses/history}.} taking the first year as the most likely; and
(iii)~a potential comet that according to England (2002) may have
been involved in a confusing story about the text on a clay tablet
found at the ancient city of Ugarit, Syria and referring to a solar
eclipse in the 14th century BC, perhaps on $-$1375~May~3. 

Hasegawa (1980) lists 11 comets chronicled by Greeks between $-$479 and
$-$336, most of them taken from Pingr\'e (1783).  In this period of
time, more Greek than Chinese comets were recorded, which is unusual.
On the one hand, Barrett (1978) cautions that some of the comets
noted by Greek historians are suspect, on the other hand Aristotle's
comet may not have been the only sungrazer observed in this period
of time.  Somewhat arbitrarily, I include the Greek comet of $-$340
--- another one depicted by Aristotle in his writings (Barrett 1978)
--- as a possible fragment of the comet of $-$1900.  I likewise
add three Chinese comets, seen according to Ho (1962) in $-$515,
$-$432, and $-$213, the last having been deemed a suspect
before (Sekanina \& Chodas 2007).  However, no Chinese comet
described as tailless ({\it po\/}) was included and neither were of
course the entries suspected of being appearances of Halley's comet.
It is rather strange that no detection of Aristotle's comet of $-$371
was noted by Chinese; Korean accounts of comet sightings seem to have
begun in the 1st century BC, Japanese not before the 7th century
AD.\,\,\,\,\,\,\,

Because of the grossly incomplete statistics of comets in the centuries
before the Common Era, it was necessary to postulate a number of missed
comets.  Fitting the pedigree chart, sungrazers were adopted to have
passed perihelion in the years $-$1300, $-$950, $-$850, $-$700,
$-$550, $-$330, and $-$260.  A total of ten missed comets had to
be introduced.

I was now ready to assemble a representative pedigree chart for a
super-Kreutz system that went back some four millennia, with the
comet of $-$1900 being~considered the progenitor or the single
sungrazer of the~first generation, with which the process of cascading
fragmentation began.  The list of the 51 potential sungrazers that
make up the chart is presented in Table~3.  Its format copies Table~1,
except that times are rounded off to the whole year.  The total
number of sungrazer generations amounts to eight; the third and
fourth generations already overlap because of massive orbital diffusion.

\begin{table*}
\vspace{0.25cm}
\hspace{0.1cm}
\centerline{
\scalebox{1}{
\includegraphics{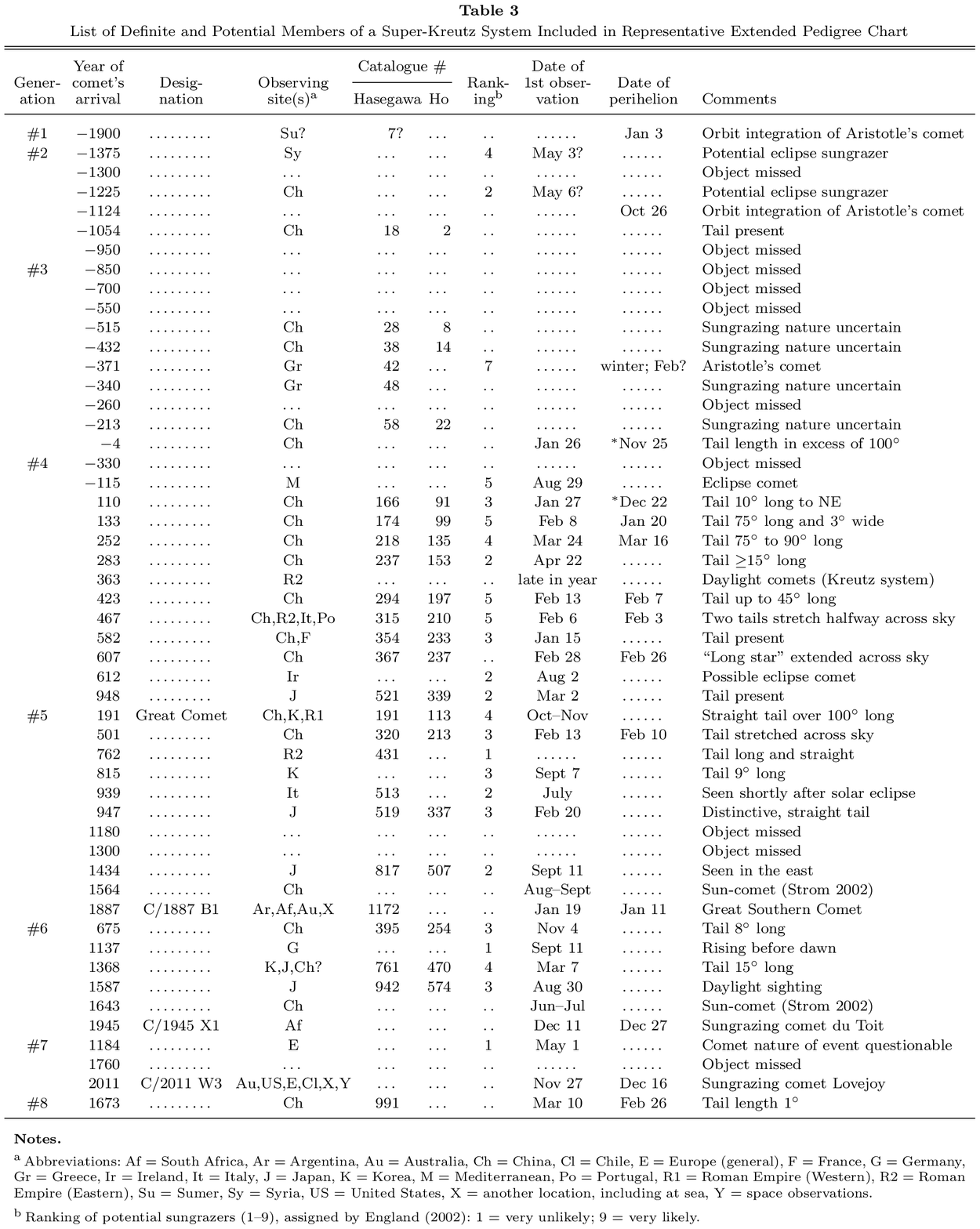}}}
\vspace{0cm}
\end{table*}
\begin{table*}
\vspace{0.25cm}
\hspace{-0.1cm}
\centerline{
\scalebox{1}{
\includegraphics{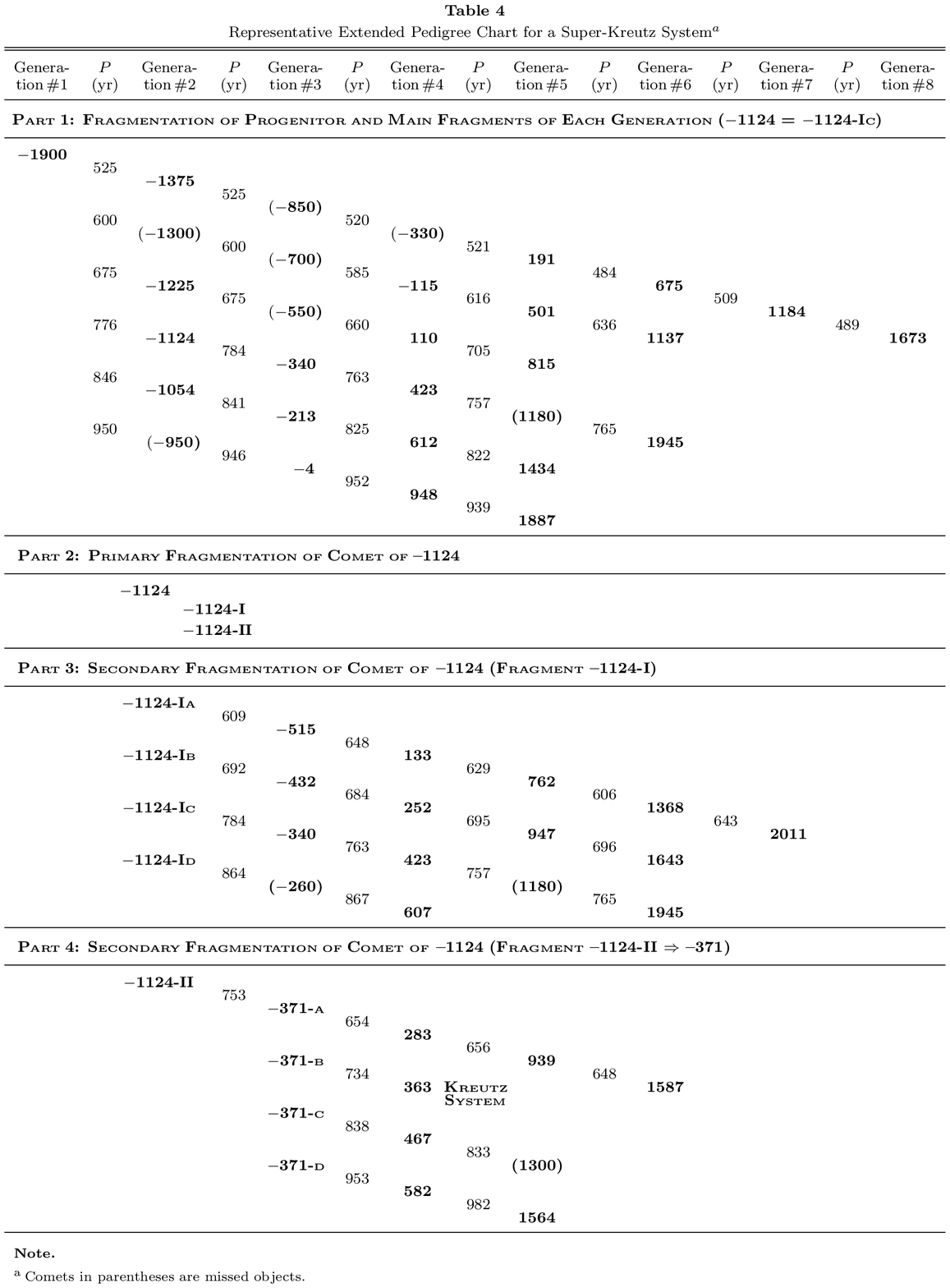}}}
\vspace{0cm}
\end{table*}

Table 4 presents the assembled pedigree chart for a super-Kreutz system.
Its format resembles the chart for the Kreutz system in Table~2, but is
substantially simplified by eliminating the columns providing $\Delta P$
and the fragment letter identification.  Table~4 consists of four parts, an
arrangement dictated by the complex structure proposed for the super-Kreutz
system.  In Part~1~I~list, from the 3rd generation on, only the so-called
{\it main fragments\/}.  A parent sungrazer's main fragment is one
characterized by the smallest shift of its center of mass from the
parent's center of mass, $\Delta U$, and therefore also by the least
magnitude of $\Delta P_{{\rm p}\rightarrow{\rm f}}$.  An example~is
provided by Table~2:\ the main fragment of the sungrazer 363-C among
the third-generation objects is the comet of 1138 (Fragment~C), whose
\mbox{$\Delta P_{{\rm p}\rightarrow{\rm f}} = +39.8$ yr} and the
main fragment of this comet among the fourth-generation objects is
the Great September Comet of 1882 (Fragment~C), whose \mbox{$\Delta
P_{{\rm p}\rightarrow{\rm f}} = -30.6$ yr}.  The fragments equivalent to
A, B, D, and E in the third generation and A, B, and D in the fourth
generation in Table~2, are not listed in Part~1 of Table~4. The
exception is the comet of $-$1900 (the progenitor) that is assumed
to have broken up at perihelion into six fragments returning between
the years $-$1375 and $-$950; the main fragment was the comet
of $-$1124 with \mbox{$\Delta P_{{\rm p}\rightarrow{\rm f}} = +8$ yr}.
The main fragments, distributed along the diagonals, are separated in
Table~4 by near-constant orbital periods, while periods of fragments
of the same generation (in columns) differ by \mbox{$\Delta P_{\rm frg}
\simeq 100$ yr}.  The fragments are typed in boldface, the orbital
periods in {\vspace{-0.1cm}}normal font.

Parts 2 to 4 of Table 4 call attention to the exceptional behavior
of the comet of $-$1124 near perihelion in comparison to the other
fragments of the comet~of~$-$1900.  The comet of $-$1124 is assumed
to have split twice in rapid succession in close proximity of
perihelion.  The primary breakup, outlined in Part~2, generated
two fragments, $-$1124-I and $-$1124-II.  The secondary breakup,
described in Part~3, involved a subsequent event, in which
$-$1124-I split into four pieces, $-$1124-I{\scriptsize A}, \ldots,
$-$1124-I{\scriptsize D}.  For each of these, a sequence of {\it main
subfragments\/} is listed.  Note that these data for
$-$1124-I{\scriptsize C}, which itself is the main fragment of the
comet of $-$1900, have to be identical to those for the comet of
$-$1124 in Part~1.

It is assumed that the evolution of the second object, $-$1124-II,
involved in the primary breakup, was very different.  It did not
fragment any longer in the year $-$1124, but split instead during
the following perihelion, in $-$371, into four pieces, which returned
in the years 283, 363, 467, and 582, respectively.  The second of the
four fragments, $-$371-{\scriptsize B}, was a contact binary, which
split near the following aphelion and then continued to fragment to
give birth to the Kreutz system proper.

\section{Implications of the Pedigree Chart\\for a Super-Kreutz System}
The 51 potential sungrazers that make up the representative pedigree
chart for a super-Kreutz system in Table~4 include 25~objects from
England's (2002) dataset, eight objects from Hasegawa-Nakano's (2001)
dataset, two sun-comets from Strom's (2002) list, 11~objects have
other sources, and 10~objects are missed comets.  Five of the 51~comets
are common to England's and Hasegawa-Nakano's datasets and six were
also listed in Table~2.

Tables 2 and 4 together comprise 20 of the 24~potential Kreutz/super-Kreutz
sungrazers (or 83~percent) tabulated by Hasegawa \& Nakano, 44 of the
62~potential Kreutz/super-Kreutz sungrazers (or 71~percent) presented
by England, and eight of Strom's 14~sun-comets (or 57~percent).
Given that the number of genuine sungrazers is likely to be at most
50~percent of the potential sungrazers among naked-eye comets (Section~3),
the pedigree charts in Tables~2 and 4 must include a number of objects
that are not sungrazers.

Two of the six entries that are in both Table~2 and Table~4 --- the
comets of $-$371 and 363 --- represent the direct link between the
Kreutz system and super-Kreutz system. Of the remaining four, three are
the known sungrazers C/1887~B1 (the Great Southern Comet of 1887),
C/1945~X1 (du Toit), and C/2011~W3 (Lovejoy), and the last one ---
the comet of 1368 --- is in the two charts directly connected to
Lovejoy (Section~2).

The 1887 sungrazer, du Toit, and Lovejoy have one major feature in
common:\ they all disintegrated in close proximity of perihelion:\ the
1887 comet about 6~hr after perihelion (Sekanina 1984) and Lovejoy about
40~hr after perihelion (Sekanina \& Chodas 2012); du Toit was never seen
(although searched for) after perihelion (Van Biesbroeck 1946, Sekanina
\& Kracht 2015).  For each of the three an evolutionary path was proposed
within the Kreutz system proper in Table~2, so the paths in Table~4 are
merely potential alternative solutions.

The disintegrating sungrazers are relevant to the issue of the lifetime
of the Kreutz/super-Kreutz system, which I briefly addressed already in
Section~3 in connection with the sungrazers' temporal distribution.  The
pedigree charts in Tables~2 and 4 show the assumed fragmentation process
to extend over as many as five and eight generations of objects,
respectively.  The data provide no information on the lifetime, as the
predicted fragmentation paths of the three disintegrating comets ---
four, five, and six revolutions about the Sun --- may or may not be
reasonable numbers.  Yet, one is probably correct in estimating that
a typical lifetime is a few revolutions and depends on the dimensions
and morphology of the fragmenting body.  Given that the origin of the
Kreutz system proper is attributed to the bifurcation of a contact
binary, it appears that the progenitor of the super-Kreutz system may
have been an {\it agglomerate of more than two\/} such parts and that
all but the last two had separated during perihelion returns prior to
the aphelion breakup.  This would mean that the {\it Kreutz system
proper could in effect represent an ultimate deagglomeration stage of
the super-Kreutz system.}

A particularly puzzling story is the orbital evolution of comet Lovejoy,
as already commented on in Section~4.  If it indeed was a fragment of the
comet of 1368, itself a member of the Kreutz system proper, then this
latter comet should belong to Population~III and in order to have reached
perihelion in AD~363, its orbital period should have been either about
500~yr or 1000~yr.  If the orbital period was $\sim$500~yr, one would
expect the comet's {\it main fragment\/} to return to perihelion during
the second half of the 19th century.  None of the known sungrazers in
that period of time appears to have been a member of Population~III,
so that the short orbital period is unlikely.  If the orbital period
was near 1000~yr, diagnostic times of the perihelion returns are a
matter of the future.  The path proposed for Lovejoy in Table~4 would
bypass a Kreutz-system scenario but leave the association with
Population~III unexplained.

Another highly controversial issue is that of the comets of 423 and
467, either of which was deemed a potential Kreutz sungrazer by
England (2002).  The two objects were independently judged attractive
Kreutz candidates in a model that was believed to be able to
simultaneously explain the sungrazer clusters in the 17th and 19th
centuries as fragmentation products of a parent, whose orbital period
was near 700~yr and which reached perihelion at the time of the
probable sungrazer X/1106~C1 (Sekanina \& Chodas 2007).  More recently,
this model was abandoned because it appeared to be inferior to the new
contact-binary model (Paper~1).  The degree of inconsistency in judging
whether the comet of 423 or 467 was more likely to be a Kreutz sungrazer
is apparent from comparison of the results of two investigations.  On the
one hand, Hasegawa \& Nakano (2001) clearly preferred the comet of 423,
as the comet of 467 did not even make their list.   On the other hand,
Mart\'{\i}nez et al.\ (2022) in their more recent extensive work reached
the opposite conclusion, agreeing that the comet of 467 could well be
a member of the Kreutz system, but arguing that the comet of 423 had a
perihelion distance of $\sim$0.47~AU.

The final problem worth commenting on in this section is the questionable
existence of potential-sungrazer pairs and/or triplets.  I leave aside
Ammianus'``daylight comets'' in AD~363 (Paper~3), the phenomenon that
is probably unique.  Rather, I refer to cases similar to the two
potential sungrazers in 1041, already mentioned in Section~4.
A difference of about two months in their arrival times at perihelion
is readily explained by the radial component of a separation velocity
of about 1~m~s$^{-1}$ if they broke up at aphelion, or 0.1~m~s$^{-1}$
if at 30~AU from the Sun on the way to aphelion.

The two comets of 1041 are by no means an isolated case of the pair
arriving in rapid succession.  Two potential sungrazers observed on,
respectively, 501 February~13 and April~14 are another example.  And
from 349/350 there are accounts of comets from December~2, January~29,
and March~30, possibly three independent objects.  While the 1041 pair
is listed as separate entries by England and in both Hasegawa's (1980)
and Ho's (1962) catalogues but not by Hasegawa \& Nakano (2001), the 501
pair is listed as two separate entries by England and by Ho but as one
in Hasegawa's catalogue.  The 349/350 triplet is archived as a single
entry by both England and Ho but in his catalogue Hasegawa lists the
March object as a separate entry.  There obviously is no consensus in
these matters.

Much confusion has surrounded the comet of 1368, suggested in Section~2
as a potential parent to the sungrazer Lovejoy of 2011.  England
(2002) considers it a Korean-Japanese comet with March~7 as the
discovery date, consistent with the reference in Ho (1962).  But
England misquotes Hasegawa (1980) about the Japanese observing the
comet the following evening.  Also, his tentative identification of
this comet with that observed by the Chinese on February~7 is in
conflict with Ho's (1962) catalogue, in which these are listed as
two different objects.  Ho identifies the comet seen in Korea during
much of April with one observed by the Chinese in early April and by
the Japanese in late April or early May.  It appears likely that two
different objects were seen between early February and mid-May.

\section{Conclusions}
%  Section 8
Once the contact-binary model is accepted, all second-generation members
of the Kreutz system (or the first-generation fragments of Aristotle's
comet via separation of the two lobes) appear as Ammianus' daylight comets
in late AD~363.  {\it All other objects\/} that were moving~in~similar
sungrazing orbits and arriving at perihelion throughout (much of) the
first millennium AD, must have been related to Aristotle's comet one
way or the other but could not have been members of the Kreutz system.
England (2002) and Hasegawa \& Nakano (2001) list, respectively, 32 and
10~candidates for such related comets.  Perfunctory examination suggests
that as many as one half of the candidates should be genuine relatives,
specifically fragments that had separated from Aristotle's comet {\it
before\/} the bond between the two lobes failed, plus subsequent
subfragments of these fragments.  In this paper I have been calling a
{\it super-Kreutz system\/} the total of this assemblage of non-Kreutz
fragments and subfragments {\it plus\/} the Kreutz system proper.
The contact binary itself may have been part of a greater
agglomerate (Aristotle's comet long before 372~BC) consisting of
morphological structures equivalent to the contact binary and all
but the two lobes may have separated at perihelion returns prior
to the aphelion breakup.  The Kreutz system could possibly represent an
ultimate deagglomeration stage of the super-Kreutz sungrazer system.

Sets of orbital elements are unavailable for a majority of potential
historical sungrazers.  Usually, the only~approximately known parameter
is the time of perihelion passage, which precedes the time of (the tail's)
peak apparent brightness by a few days to a few weeks at most.  To assemble a
pedigree chart, one needs to apply a plausible algorithm that is supported
by data and is dynamically justified.  I use a steplike increase in the
orbital period of fragments, which seems to be closely followed by the
tidally-driven post-perihelion motions of the secondary nuclei of the Great
September Comet of 1882 and is fairly consistent with a constraint provided
by an investigation of Roche's limit.  Whereas tidally produced fragments
move in orbits of vastly different periods and are subjected to rapid
diffusion, bumps on a cumulative distribution curve of potential historical
sungrazers show that fragmentation also proceeds at large heliocentric
distance.  As expected, over short time periods this process increases the
number of sungrazing comets, but eventually, over longer periods, the
disintegration effect prevails and their number declines.

The fragmentation paths in the representative pedigree charts for both
the Kreutz system and a super-Kreutz system should not be taken literally.
In the displayed exercise they are merely examples that conform to
very limited constraints, but apart from that they are arbitrary, as is
the choice of missed objects.  Only the few links characteristic of the
contact-binary model are meaningful to the extent the
model is.  Evidence for a super-Kreutz sungrazer system of young age,
incorporating the Kreutz system proper as its part, is overwhelming and
the selected pedigree charts are in line with this evidence.

\vspace{0.2cm}
\begin{center}
{\footnotesize REFERENCES}
\end{center}

\vspace{-0.35cm}
\begin{description}
{\footnotesize
\item[\hspace{-0.3cm}]
Abney,\,W.\,de\,W.,\,\&\,Schuster,\,A.\,1884,\,Phil\mbox{.\hspace{0.007cm}T}rans\mbox{.\hspace{0.02cm}R}oy.\,Soc.\,London,{\linebreak}
 {\hspace*{-0.6cm}}175, 253
\\[-0.57cm]
\item[\hspace{-0.3cm}]
Aggarwal, H.\ R., \& Oberbeck, V.\ R.\ 1974, ApJ, 191, 577
\\[-0.57cm]
\item[\hspace{-0.3cm}]
Barrett, A.\ A.\ 1978, J.\ Roy.\ Astron.\ Soc.\ Canada, 72, 81
\\[-0.57cm]
\item[\hspace{-0.3cm}]
Bortle, J.\ E.\ 1998, {\tt http://www.icq.eps.harvard.edu/bortle.html}
\\[-0.57cm]
\item[\hspace{-0.3cm}]
England, K.\ J.\ 2002, J. Brit.\ Astron.\ Assoc., 112, 13
\\[-0.57cm]
\item[\hspace{-0.3cm}]
Green, D.\ W.\ E., ed.\ 2020, {\tt http://www.icq.eps.harvard.edu/}{\linebreak}
 {\hspace*{-0.6cm}}{\tt brightest.html}
\\[-0.57cm]
\item[\hspace{-0.3cm}]
Hasegawa, I.\ 1980, Vist.\ Astron., 24, 59
\\[-0.57cm]
\item[\hspace{-0.3cm}]
Hasegawa, I., \& Nakano, S.\ 2001, Publ.\,Astron.\,Soc.\,Japan, 53, 931
\\[-0.57cm]
\item[\hspace{-0.3cm}]
Ho, P.-Y.\ 1962, Vist.\ Astron., 5, 127
\\[-0.57cm]
\item[\hspace{-0.3cm}]
Kreutz, H.\ 1891, Publ.\ Sternw.\ Kiel No.\ 6
\\[-0.57cm]
\item[\hspace{-0.3cm}]
Marsden, B.\ G.\ 1967, AJ, 72, 1170
\\[-0.57cm]
\item[\hspace{-0.3cm}]
Marsden, B.\ G.\ 1989, AJ, 98, 2306
\\[-0.57cm]
\item[\hspace{-0.3cm}]
Mart\'{\i}nez, M.\,J., Marco, F.\,J., Sicoli, P., \& Gorelli, R.\ 2022,
 Icarus,{\linebreak}
 {\hspace*{-0.6cm}}384, 115112
\\[-0.57cm]
\item[\hspace{-0.3cm}]
Pingr\'e,\,A.\,G.\,1783, Com\'etographie ou\,Trait\'e historique et\,th\'eorique{\linebreak}
{\hspace*{-0.6cm}}des com\`etes. Tome Premier. Paris: Imprimerie Royale
\\[-0.57cm]
\item[\hspace{-0.3cm}]
Rolfe, J.\ C.\ 1940, The Roman History of Ammianus Marcellinus,{\linebreak}
 {\hspace*{-0.6cm}}{Book XXV. {\tt
 https://penelope.uchicago.edu/Thayer/E/Roman/}}{\linebreak}
 {\hspace*{-0.6cm}}{\tt Texts/Ammian/25$^\ast\!\!$.html}
\\[-0.57cm]
\item[\hspace{-0.3cm}]
Schmitt, A. 1949, IAU Circ.\ 1221
\\[-0.57cm]
\item[\hspace{-0.3cm}]
Sekanina, Z.\ 1984, Icarus, 58, 81
\\[-0.57cm]
\item[\hspace{-0.3cm}]
Sekanina, Z.\ 2021, eprint arXiv:2109.01297 (Paper 1)
\\[-0.57cm]
\item[\hspace{-0.3cm}]
Sekanina, Z.\ 2022a, eprint arXiv:2211.03271 (Paper 2)
\\[-0.57cm]
\item[\hspace{-0.3cm}]
Sekanina, Z.\ 2022b, eprint arXiv:2202.01164 (Paper 3)
\\[-0.57cm]
\item[\hspace{-0.3cm}]
Sekanina, Z.\ 2022c, eprint arXiv:2212.11919 (Paper 5)
\\[-0.57cm]
\item[\hspace{-0.3cm}]
Sekanina, Z., \& Chodas, P.\ W.\ 2007, ApJ, 663, 657
\\[-0.57cm]
\item[\hspace{-0.3cm}]
Sekanina, Z., \& Chodas, P.\ W.\ 2008, ApJ, 687, 1415
\\[-0.57cm]
\item[\hspace{-0.3cm}]
Sekanina, Z., \& Chodas, P.\ W.\ 2012, ApJ, 757, 127
\\[-0.57cm]
\item[\hspace{-0.3cm}]
Sekanina, Z., \& Kracht, R.\ 2015, ApJ, 815, 52
\\[-0.57cm]
\item[\hspace{-0.3cm}]
Sekanina, Z., \& Kracht, R.\ 2022, eprint arXiv:2206.10827\,(Paper~4)
\\[-0.57cm]
\item[\hspace{-0.3cm}]
Strom, R.\ 2002, A\&A, 387, L17
\\[-0.64cm]
\item[\hspace{-0.3cm}]
Van Biesbroeck, G.\ 1946, Pop.\ Astron., 54, 154}
\\[-0.73cm]
% \vspace{-0.39cm}
%
\end{description}
\end{document}